\documentclass[conference]{IEEEtran}
\IEEEoverridecommandlockouts
\usepackage{cite}
\usepackage{amsmath,amssymb,amsfonts}
\usepackage{graphicx}
\usepackage{textcomp}
\usepackage{xcolor}
\usepackage{caption}
\usepackage{subcaption}
\usepackage{listings}
\usepackage[linesnumbered,ruled,vlined]{algorithm2e}
\usepackage[normalem]{ulem}
\usepackage{url}

\def\BibTeX{{\rm B\kern-.05em{\sc i\kern-.025em b}\kern-.08em
    T\kern-.1667em\lower.7ex\hbox{E}\kern-.125emX}}
\begin{document}

\title{GPU-RANC: A CUDA Accelerated Simulation Framework for Neuromorphic Architectures}

\makeatletter
\newcommand{\linebreakand}{%
  \end{@IEEEauthorhalign}
  \hfill\mbox{}\par
  \mbox{}\hfill\begin{@IEEEauthorhalign}
}
\makeatother

\author{
\IEEEauthorblockN{Sahil Hassan, Michael Inouye, Miguel C. Gonzalez, Ilkin Aliyev, Joshua Mack, Maisha Hafiz, Ali Akoglu}
\IEEEauthorblockA{\{sahilhassan, mikesinouye, migor2d2, ilkina, jmack2545, mhafiz1, akoglu\}@arizona.edu}
\textit{%
  Department of Electrical \& Computer Engineering, University of Arizona, Tucson, AZ}
}

\maketitle
\begin{abstract}
Open-source simulation tools play a crucial role for neuromorphic application engineers and hardware architects to investigate performance bottlenecks and explore design optimizations before committing to silicon. Reconfigurable Architecture for Neuromorphic Computing (RANC) is one such tool that offers ability to execute pre-trained Spiking Neural Network (SNN) models within a unified ecosystem through both software-based simulation and FPGA-based emulation. RANC has been utilized by the community with its flexible and highly parameterized design to study implementation bottlenecks, tune architectural parameters or modify neuron behavior based on application insights and study the trade space on hardware performance and network accuracy.  In designing architectures for use in neuromorphic computing, there are an incredibly large number of configuration parameters such as number and precision of weights per neuron, neuron and axon counts per core, network topology, and neuron behavior. To accelerate such studies and provide users with a streamlined productive design space exploration, in this paper we introduce the GPU-based implementation of RANC. We summarize our parallelization approach and quantify the speedup gains achieved with GPU-based tick-accurate simulations across various use cases. We demonstrate up to 780 times speedup compared to serial version of the RANC simulator based on a 512 neuromorphic core MNIST inference application. We believe that the RANC ecosystem now provides a much more feasible avenue in the research of exploring different optimizations for accelerating SNNs and performing richer studies by enabling rapid convergence to optimized neuromorphic architectures.
\end{abstract}  


\section{Introduction}
Neuromorphic computing architectures have emerged as a viable alternative to the Von Neumann based systems delivering significantly greater energy efficiency through event-driven execution.  
Initially designed for neural-inspired computations targeting sensing, detection and classification types of applications~\cite{IJCNN_2013_Applications, ICRC_2016_Applications}, foundational features of neuromorphic systems are motivating studies to expand the scope of neuromorphic computing into diverse application domains by leveraging energy efficient computation and exploiting high degree of concurrency scaling to millions of neuron level computations. Coupling the memory-integrated computing and fine-grained parallelism with intra-chip scalability has opened pathways for researchers to target non-machine learning applications~\cite{aimone_2022,Christensen22roadmap, davies2021advancing, Schuman22opportunity} that can be decomposed to parallel spike-based computations in the domains such as graph algorithms~\cite{schuman2019shortest,kay2020neuromorphic}, constraint optimization~\cite{fonseca2017using}, signal processing~\cite{Orchard2021signal}, and error correction~\cite{hassan_2023_RANC-LDPC}. Breaking down an algorithm into spike-based computations poses a notable challenge, but an even more significant challenge is the ability to match these computations architecturally for a resource-efficient implementation without sacrificing from energy efficient even driven execution. Achieving this demands a simulation environment that provides the flexibility to adjust architectural features at the neuron block, core, and network topology levels. This capability empowers hardware architects and application engineers to explore and fine-tune parameters of their neuromorphic architecture. Such adjustments are often impractical on a purely prefabricated ASIC or without incurring the expenses associated with custom silicon~\cite{akopyan_2015_truenorth, davies_2018_loihi, gonzalez_2024_spinnaker2}.

In an effort to bridge these gaps, we have recently introduced the Reconfigurable Architecture for Neuromorphic Computing (RANC) ~\cite{valancius_2020_fpga-emulation, mack_2021_RANC} as an open-source ecosystem \footnote{Available at \url{https://ua-rcl.github.io/projects/ranc/}}. RANC has been developed as a highly flexible environment to make neuromorphic architectural design accessible to a variety of researchers and application developers. RANC facilitates rapid experimentation with neuromorphic architectures within a unified ecosystem available in both software through Python, C++ simulation and hardware through FPGA-based emulation. RANC supports the execution of pre-trained Spiking Neural Network (SNN)  models on a multi-core architecture similar to the TrueNorth and Loihi 2. Due to its reconfigurability, RANC has been utilized by the community~\cite{clair_2023_spikehard, nguyen_2022_sleepingposture} as a baseline design to tune architectural parameters and explore different optimizations for accelerating SNNs. Consequently, RANC has emerged as a suitable candidate for integration into a heterogeneous System-on-Chip (SoC)~\cite{clair_2023_spikehard}.

Conducting experiments, such as test set inference for classification networks or vector matrix multiplication (VMM) on RANC, results in a substantial amount of simulation time, scaling to the order of hours. The duration of a simulation largely depends on the configuration of system parameters and input size as RANC offers ability to configure architectural parameters, including crossbar configuration in terms of the number of axons and neurons, the number of weights per neuron, and the bitwidths of weight, leak, neuron potential, and reset potential values. Performing a comprehensive sweeping experiment across these parameters to study the tradeoffs between energy efficiency and throughput can extend simulations to the several weeks timescale. Additionally, each incremental hardware modification necessitates thorough validation through software-based simulation using a suite of test cases before committing to FPGA-based emulation studies. In order to accelerate such studies and provide users with a streamlined productive design space exploration, we introduce the GPU-based implementation of RANC in this work~\footnote{Available at \url{https://github.com/UA-RCL/RANC/tree/gpu-ranc}}. We summarize our parallelization approach and quantify the speedup gains achieved with GPU-based simulations across various use cases.  
The significant acceleration offered by GPU-RANC allows us to conduct performance evaluation of SNNs over a much wider hyperparameter space. Moreover, it combines the ease of novel architectural modifications with detailed evaluation capability, while reducing the simulation time. In addition, verification and profiling of novel non-cognitive applications on neuromorphic platforms require detailed simulation covering wide input scenarios over longer simulation times. As the complexity of applications and input sizes increases, simulations of this nature can seamlessly extend from hours to months in scale~\cite{hassan_2023_RANC-LDPC}. Hence, GPU-RANC offers a viable approach in exploring non-cognitive application mapping within research.

Rest of the paper is organized as follows. In Section~\ref{sec:background}, we summarize the RANC simulation environment and its core components. We then present parallelization approach taken when implementing each of the components on the GPU in Section~\ref{sec:methodology}. We present our validation approach for the GPU-based implementation and quantify the speedup gains with respect to serial version of the RANC in Section~\ref{sec:results}. Finally we present our conclusions and future work in Section~\ref{sec:conclusions}.

\begin{figure}[t]
    \centering
    \includegraphics[width=\linewidth]{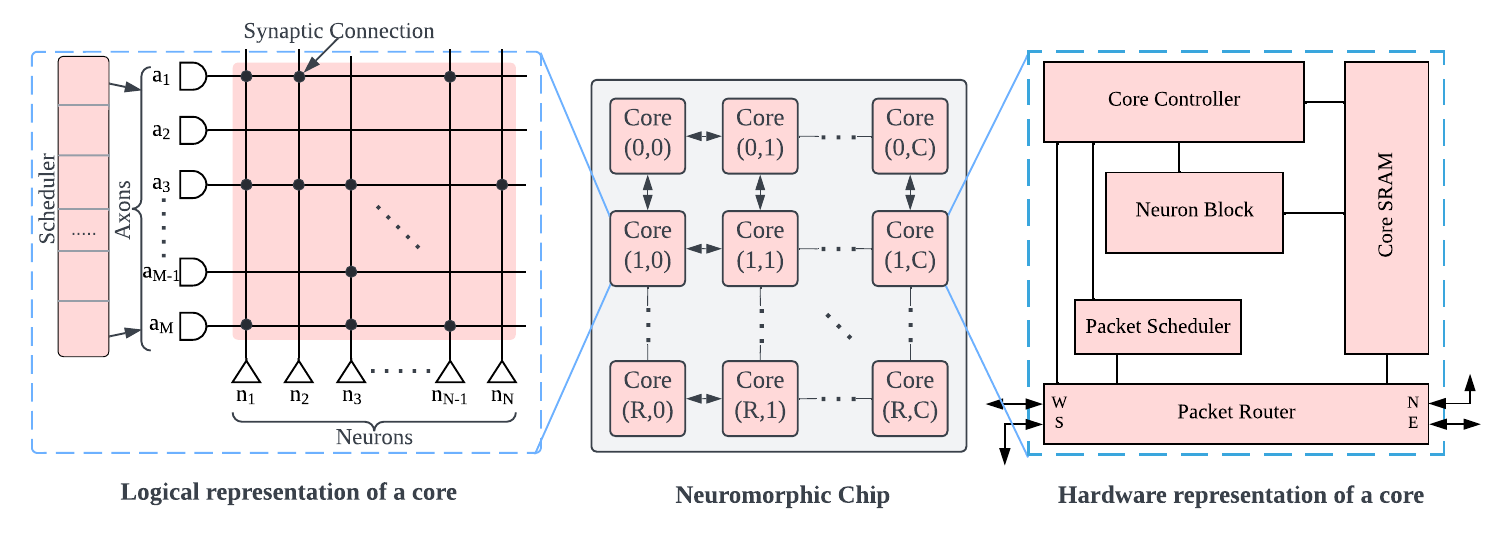}
    \caption{Overview of the RANC architecture.}
    \vspace{-4mm}
    \label{fig:RANC_overview}
\end{figure}

\section{BACKGROUND}
\label{sec:background}

RANC simulates neuromorphic cores organized in a 2D mesh-based grid, where each core can communicate directly with its  neighbouring cores (east, west, north and south directions), as shown in Figure~\ref{fig:RANC_overview}.
Each core is composed of compute and memory components to simulate the behavior of \emph{Axons}, \emph{Neurons}, \emph{Core SRAM (CSRAM)}, \emph{Scheduler},  \emph{Router}, and \emph{Core Controller}. The \emph{Axons} are an array of input nodes, where each axon receives a single bit (called a spike, with possible values of 0 or 1) of input. The \emph{Neurons} (corresponding to \emph{Neuron Block} in hardware representation) are an array of primary computation units responsible for calculating the neuron potential, and generating spikes as output of the core. 
The cross points where an axon and neuron meet are called synapses. These synapses use a configurable parameter called Synaptic Connections, which specifies whether the neuron and axon corresponding to the particular synapse are connected. 
Input received by an axon is considered in a neuron's computation only if there is a synaptic connection between that axon and neuron.
Each axon within a core has an associated weight determined by the axon type at initialization and organized as sets of four weights per neuron, known as a synaptic weight. Each neuron computation produces a potential value as well as spike outputs using the LIF neuron model.

The \emph{CSRAM} is the memory component that holds parameters associated with each neuron. These parameters are- synaptic weights, synaptic connections, and neuron potential. The \emph{Router} is the component of a core that is responsible for packaging the spikes generated in the core as a packets, and routing it to the desired destination core. The \emph{Scheduler} receives the input spike packets and sends it to the intended axon at the desired timestep. The \emph{Core Controller}  initiates spike processing, maintains/orchestrates Neuron Block operation and CSRAM data access, and maintains correct data flow between all of the core components.
The overall RANC components are synchronized by timestep known as a \emph{tick}. At each tick, all the core components perform a specific set of operations, and once all the components are done, the execution proceeds to the next tick.

\begin{algorithm}[t]
\caption{RANC Simulator Algorithm}
\label{alg:ranc_algorithm}
    Initial setup and input decode\\
    \For{tick in num\_ticks}
        {
        \For {core in num\_cores}
            { 
            clear old contents of scheduler SRAM\\
            shift in current tick input to scheduler SRAM
            }

        \For{ packet in num\_packets[tick]}
            {
            calculate destination / offset\\
            route packet\\
            write into SRAM\\
            }
        \For{core in num\_cores}
            {
            \For{neuron in num\_neurons}
            {
                \For{axon in num\_axons}
                {
                    accumulate neuron potential\\
                }
            additional computation / checks\\
            }
            }
        \For{core in num\_cores}
            {
            \For{neuron in num\_neurons}
                {
                \If{neuron$\rightarrow$ spike == 1}
                    {
                        calculate destination / offset\\
                        route packet\\
                        write into SRAM\\
                    }
                }
            }
        }
\end{algorithm}

The execution flow of RANC software simulator is captured by Algorithm~\ref{alg:ranc_algorithm}. Here, the simulator starts by initializing RANC configuration and receiving input data. Therefore, the \texttt{for} loop on line 2 starts the simulation for the number of ticks specified by user. Within each tick's execution, the simulator first iterates over all cores (line 3) in order to clear the backdated data from their schedulers and schedule the updated input data to be sent to the axons. Once this process is done, the simulator iterates over all the packets scheduled for the current tick (line 6), and calculates their destination core, routes the packets, and writes them onto the Scheduler SRAM of cores. The simulator starts the core-level execution process by iterating over each core (line 10). Within each core, the simulator iterates over each neuron (line 11), and for each neuron it iterates over all relevant axons to accumulate the product of their respective spikes and weights in the neuron potential sum (lines 12-13). Additional operations such as applying leak value, comparing threshold to generate spike and resetting neuron potential are performed at line 14. Finally, simulator iterates over each core (line 15) and each neuron within cores (line 16) to identify neurons that produce spike. The simulator packs spikes into packets, routes them to their desired cores and writes into the core SRAMs.

\section{METHODOLOGY}
\label{sec:methodology}
Algorithm~\ref{alg:ranc_algorithm} shows the tick-by-tick simulation flow in RANC that involves dedicated loops for the Scheduler (lines 3-5), Router (lines 6-9 and 15-20)) and Neuron Block (lines 10-14) operations. Profiling the workflow in RANC shows that  
The Scheduler, Router, and Neuron Block loops take 2.2\%, 5.8\% and 91.7\% of the total simulation time respectively. Therefore our parallelization focus will be on the Neuron Block. However, both Scheduler and Router blocks need to be unrolled to realize tick-level concurrent execution flow as we unroll the triple nested for loop of the Neuron Block. Furthermore, unrolling these three blocks also require replication type of modifications to the Core Controller and CSRAM implementations to support concurrency and launch all neuromorphic cores in parallel at the grid level on the GPU.
Grid level parallel execution can have potential savings for all kernels since data transfer overhead caused by the host-GPU communication is often a significant performance bottleneck. Across this work, we will associate the versions of our kernels to a grid-level or core-level optimization.

\subsection{Core Controller}
The neuron block datapath operations and memory transactions are governed by the Core Controller implemented with a state machine. Therefore, there is no parallelism opportunity within each core, as certain behavior within a core is invoked based on the order dictated by the state machine. The simulation flow iterates through each core sequentially, and then executes its high-level control logic. In hardware implementation, all cores across the RANC grid execute concurrently within a tick. When constructing kernels for the neuron block, router and scheduler blocks, we integrate the matching core controller logic into each respective kernel so that the entire grid can be updated with maximum concurrency.

\subsection{Core SRAM}
\label{sec:coresram}
The Core SRAM (CSRAM) contains parameters to define a single neuron, such as current potential, weights, and connections, and these parameters are used in both neuron block and router operations. Due to its non-contiguous memory layout, the CSRAM is not eligible for GPU acceleration. We modify the data structure to be a contiguous global array accessible by the neurons of the entire core. This modification allows CSRAM to be allocated and copied to the GPU global memory once to avoid core level repetitive copies.

\subsection{Neuron Block}

Accumulations performed on neuron potential are independent of each other at neuron block level during each tick as illustrated in Algorithm~\ref{alg:ranc_algorithm} lines 10-14. We implement neuron block level optimizations in three steps as summarized below.

We launch all neurons of a single core in parallel (line 11), and computations for each neuron within the core are mapped to an individual GPU thread (\emph{Core-level optimization}). The inputs such as, weights, synaptic connections, input spikes at axons are all maintained in the global memory of the GPU.
Each thread executes the \texttt{for} loop (lines 12-13) to calculate the neuron potential by iterating over all the axons and accumulating the weight associated with the axons given there exists a synaptic connection and the axon has received an input spike. Upon calculating the potential, each thread checks if the current neuron potential exceeds the threshold and generates a spike if the condition is satisfied. 

We then expand neuron block level parallelism to grid level (lines 10-11) where a single kernel launch on GPU consists of all neurons belonging to all cores of the RANC 2D grid at each tick (\emph{Grid-level optimization}). Implementation involves changing the data structure of RANC cores and its underlying components from vectors and objects to GPU-friendly flattened arrays. Realizing this level of concurrency is limited by the threading capacity of the target GPU.   

Finally, the loop at line 13 of Algorithm~\ref{alg:ranc_algorithm} represents axon accumulation, which is suitable for reduction tree based implementation. Therefore, we change the thread to data mapping from ``thread to neuron" to ``thread to axon" (\emph{Synapse-level optimization}). Along with this, we copy synaptic weights to shared memory for faster access, adjust the reduction tree to process in a reverse pattern to avoid bank conflicts and thread divergence and, use the initial shared memory load phase as first stage of the reduction to minimize the thread workload.

\subsection{Router}
As neurons generate spikes, each spike is sent to local router. The Router Block generates a packet for each spike and packets are transferred to the destination core by hopping through the router components of neighbouring cores. At the destination core, the packet writes its target tick offset and destination axon location into the SRAM.
The serial RANC simulator routes each spike packet a single step at a time in either the east, west, north or south direction sequentially until the packet reaches its destination core. 
As illustrated in Figure \ref{fig:router_metho_cpu}, assuming $(n, m)$ is the grid dimension, a packet will take $n+m-2$ steps to reach core $(n-1, m-1)$ from core $(0, 0)$ for the worst case scenario. Since we are not limited by hardware behavior and the packet already contains information to determine the final destination core, we calculate the packet's destination core directly and route the packet in a single step. 
We assign a GPU thread for each packet as all packets are independent of each other. The router kernel makes use of the \emph{CSRAM} and output spikes loaded into the global memory of the GPU. In order to amortize the data transfer overhead between host and GPU, we launch the kernel at a grid level, where all the spikes generated at a certain tick gets processed in parallel.
We modify the data structure of all scheduler SRAMs from C++ class members to a single array, which is sent once to the GPU at the start of the RANC simulation as described in Section \ref{sec:coresram}. When a packet needs to be sent to another core due to a spike, a spike bit is written directly into the scheduler SRAM array by the GPU thread.

\begin{figure}[t]
\centering
    \begin{subfigure}[b]{.22\textwidth}
        \centering
         \includegraphics[width=0.75\linewidth]{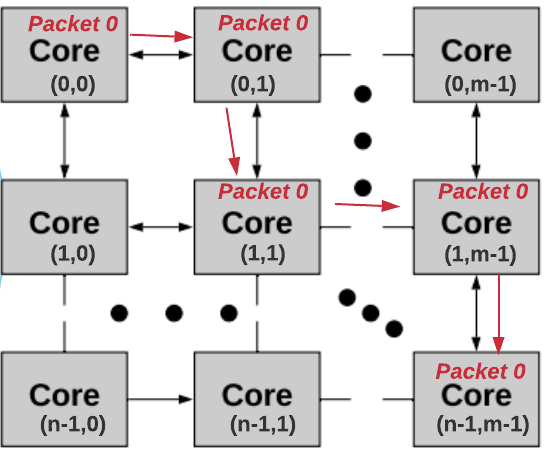}
         \caption{Original C++ Mapping}
         \label{fig:router_metho_cpu}
    \end{subfigure}
    \begin{subfigure}[b]{.22\textwidth}
        \centering
         \includegraphics[width=0.75\linewidth]{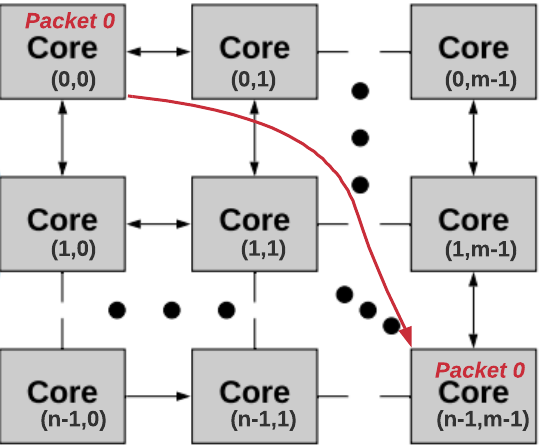}
         \caption{Optimized CUDA Mapping}
         \label{fig:router_metho_gpu}
    \end{subfigure}
\caption{Router mapping illustration where packets move across each core in the reference serial implementation and packets are routed to destination directly to the destination to match the concurrency level of the neuron block.}
\vspace{-4mm}
\end{figure}

\subsection{Scheduler}
\label{sec:methodology_scheduler}

A scheduler object within the serial simulator is instantiated for each core, containing a 2D array of booleans with a length equal to the maximum tick offset and a width equal to the number of axons. This array stores input spikes for its respective core's axons from the current tick up to the maximum tick offset into the future. 
At the beginning of each tick, spike values are cleared out from the index of the last tick's axon array as they are now obsolete as depicted in Figure \ref{fig:schedulerc}. This frees up an additional array of possible spikes at a maximum offset into the future. Then, the index that the scheduler uses to access the current tick's spikes is incremented or reset if it has reached the maximum value. 
In the serial flow, schedulers are iterated through sequentially from core $0$ to core $N$ and then data is set one element at a time. The step complexity of the existing algorithm is therefore $(N \times num\_axons)$.
As individual axons are independent of one another, we assign each individual axon data manipulation to a GPU thread as shown in Figure \ref{fig:schedulercuda}. The full contents of the original scheduler data array for every core is combined into a single 3D structure, first indexed by core, and then by tick offset and axon number as before. Similar to the optimizations made for the CSRAM data object, we condense the cores' discrete arrays to limit the number of distinct CUDA memory allocations and copies to the device global memory. This allows a single kernel to have full access to all current and future spikes throughout the device, allowing a thread to write into the correct axon regardless of its assigned core. After the completion of coalesced writes, a single thread per core must increment the index pointing to the spikes for current tick ($curr\_word\_index$). Instead of choosing only the first thread in each block to accommodate this, we pack all of the operations into the first few thread blocks by checking the global thread index to minimize thread divergence. Without any coupling between core axon count and the CUDA kernel block size, we run a sweeping experiment through different possible block configurations to determine an optimal setup. 

    \begin{figure}[t]
    \centering
    \begin{subfigure}[b]{.36\textwidth}
         \centering
         \includegraphics[width=.85\linewidth, trim={5cm 5.5cm 6cm 4cm}, clip=true]{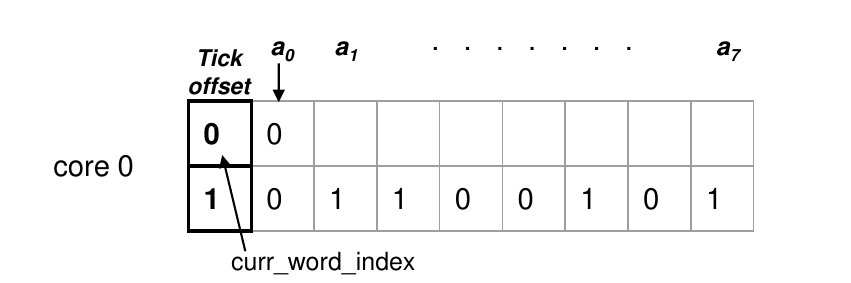} 
         \caption{Original C++ Mapping}
         \vspace{0.1cm}
         \label{fig:schedulerc}
    \end{subfigure}
    \hfill
    \hfill
    \begin{subfigure}[b]{.36\textwidth}
         \centering
         \includegraphics[width=.82\linewidth, trim={5cm 3cm 6.5cm 3cm}, clip=true] {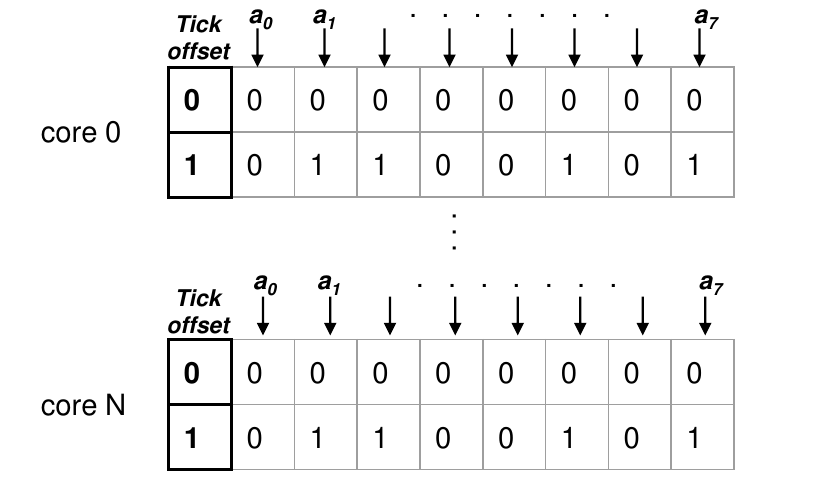} 
         \caption{Optimized CUDA Mapping}
         \label{fig:schedulercuda}
    \end{subfigure}
    \caption{Scheduler mapping where max tick offset is two and there are eight axons ($a_0$ through $a_7$) per core.}
    \vspace{-12pt}
    \label{fig:schedulermapping}
\end{figure}

\section{EXPERIMENTAL RESULTS}
\label{sec:results}

A suite of different applications are used to functionally validate and characterize the performance of our CUDA implementation. Among them are three different SNNs to perform inference of the MNIST dataset \cite{lecun-mnisthandwrittendigit-2010} with 12, 128, and 512 RANC core counts, vector matrix multiplications (VMMs) of two 32x32, 50x50, and 60x60 matrices, and an SNN to perform classification of the CIFAR-10 data set \cite{Krizhevsky2009LearningML}. The varying core counts within the MNIST runs represent that the networks trained off-chip have varying numbers of parameters, with higher core counts usually used for higher classification accuracy. 
The simulation time of these test applications on the serial CPU-based RANC ranges from around 65 seconds (VMM 32x32) to over 5.6 hours (MNIST with 512 RANC cores).
In addition to these, we also simulate a large RANC grid with no application mapped that still performs all necessary actions for a tick update but with no spikes generated within the system. 
When choosing grid parameters for this scenario, the system is configured to resemble IBM's TrueNorth neuromorphic ASIC. Hence we call this design as TrueNorth Reference. The intention of this test case is to gain insight on general performance trends as we scale the RANC parameters to the extreme. All specific application configuration parameters are presented in Table \ref{tab:applications}.

\begin{table}[t]
    \centering
    \caption{Applications and test parameters.}
    \begin{tabular}{|c|c c c c c|} 
        \hline
        Application & Ticks & \# of & Axons & Neurons & CPU exec.\\
         & & Cores & /Core & /Core & time (s)\\
        \hline
        MNIST-12c & 10010 & 12 & 256 & 256 & 539\\ 
        \hline
        MNIST-128c & 10010 & 128 & 256 & 256 &  5230\\ 
        \hline
        MNIST-512c & 10010 & 512 & 256 & 256 & 20344\\  
        \hline
        VMM 32x32 & 788 & 21 & 256 & 256 & 65\\ 
        \hline
        VMM 50x50 & 889 & 45 & 512 & 256 & 311\\ 
        \hline
        VMM 60x60 & 1095 & 51 & 512 & 256 & 434\\
        \hline
        CIFAR10 & 10010 & 36 & 256 & 256 & 1693\\ 
        \hline
        TrueNorth Ref. & 500 & 4096 & 256 & 256 & 8295\\ 
        \hline
    \end{tabular}
    \label{tab:applications}
     \vspace{-3mm}
\end{table}

A test script is created that simulates each of the applications on both the original CPU-based C++ version and GPU-based CUDA version along with the addition of some execution time profiling code which allows us to compare kernel-specific performance. The addition of profiling code adds a minimal amount of overhead, but we believe that it makes up a negligible percentage of the total simulation time. Each application generates an output file containing spikes from the output bus that we can use to functionally validate the GPU accelerated implementation. RANC contains no stochastic effects, so a one-to-one match between output files across all applications from the CPU and GPU-based versions gives us confidence of equivalence. For both versions, all operations or kernels including data copy overhead is also recorded. 

Our reference system that we run all simulations on contains an AMD EPYC 7552 48-core processor clocked at 2.9GHz and an NVIDIA Tesla V100S 32GB. Neither the existing C++ simulator nor our CUDA implementation uses any multi-threading, so the 48 core system does not offer any added benefit to either simulator's performance. 
In the following subsections, kernel-specific results and trends are first discussed followed by an overall simulation time analysis. 

\subsection{Neuron Block}

In order to quantify the speedup gained by the neuron block optimizations, as well as the variation in speedup due to varying RANC core counts, we run RANC simulations with MNIST applications for core counts of 12, 128 and 512 with GPU block size of 1024. The selection of this block size is made in order to maximize the thread utilization in 1D without running into any memory limitations. After running these experiments, we calculate the speedup gained by neuron block optimizations compared to CPU-based RANC. We run these experiments using GPU-RANC with core-level, grid-level and synapse-level neuron block optimizations.

Figure~\ref{fig:nb_all_result} presents speedup gained by every neuron block optimization compared to CPU-based RANC. The X-axis shows the varying core count and Y-axis represents the speedup achieved compared to the CPU-based RANC. 
From the core-level optimization trend shown by the red line, we observe a speedup gain of 113x for MNIST-12c. As the application size grows to 128 cores (MNIST-128c), the speedup gain increases to 137x. However, for further growth in core count (MNIST-512c) the speedup gain saturates.
This indicates that core-level optimization is not able to exploit the parallelism potential.

\begin{figure}[t]
    \centering
    \includegraphics[width=.9\linewidth]{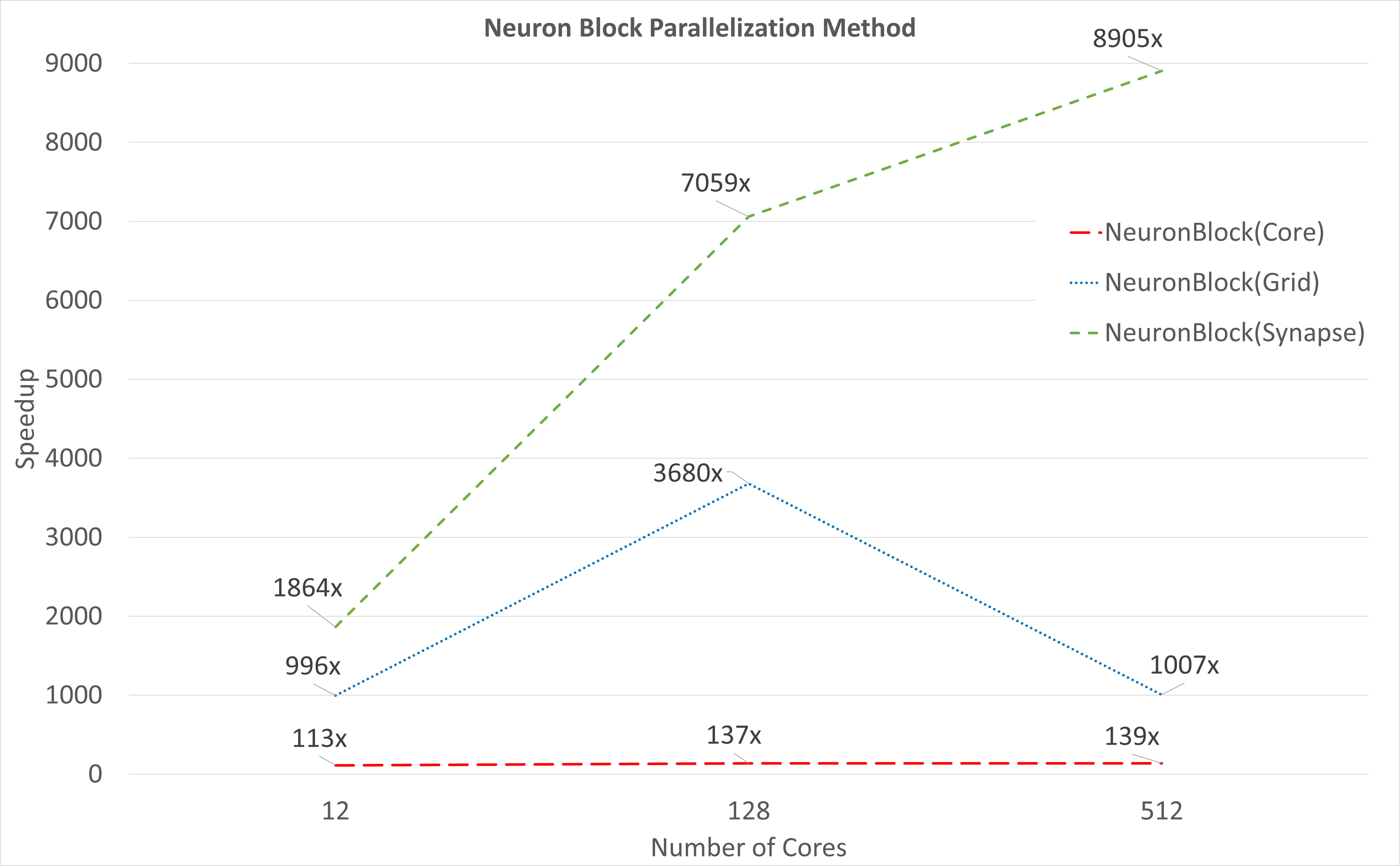}
    \caption{Speedup by different levels of Neuron Block optimizations across core counts over serial execution using MNIST.}
     \vspace{-4mm}
    \label{fig:nb_all_result}
\end{figure}

Focusing on the yellow trend line for grid-level optimization, we observe a kernel speedup ranging from 996x to 3680x compared to serial execution, which is a significant improvement over core-level optimization.
Following the trend of speedup with increasing core count, we notice that as the core count increases from 12 to 128, speedup gain jumps from 996x to 3680x. This gain is obvious as for larger number of cores, more neurons are launched in parallel. However, as core count increases further to 512, the speedup gain drops from 3680x to 1007x. This behaviour, although unexpected, can be explained by the GPU thread launch limit. The V100 has 56 Streaming Multiprocessors with 2048 threads each, resulting into a total of 115,000 threads. However, launching 512 cores requires 131,072 threads (512 cores $\times$ 256 neurons), which is more than the available threads. The scheduler within the GPU therefore serializes this kernel launch to accommodate the computations, resulting in reduced speedup. 

Finally, based on the performance trend of synapse-level optimization shown with the green line, we observe the largest speedup gain in the range of 1864x to 8905x. This large gain can be attributed to the fine-grained parallelism achieved by unrolling all of the loops during neuron computations, as well as the use of shared memory for faster data access while avoiding bank conflict and thread divergence. Looking at the trend in speedup with respect to increase in core count, we note that the speedup achieved by moving from 12 to 128 cores (5195x), is much larger than the speedup gained by moving from 128 to 512 cores (1846x). This indicates a saturation trend for core counts larger than 512.

\subsection{Router and Scheduler}
\label{router_results}
Even though the neuron block operations dominate the total execution time in the RANC simulator, we choose to implement the router and scheduler components on the GPU as well. In this section we present thread block configuration analysis we performed for these two kernels for reproducibility of our results. We evaluate the execution time performance of the router kernel referring to Figure~\ref{fig:routerblocksize}, where X-axis represents varying thread block size and Y-axis represents the speedup compared to the CPU-based RANC. We observe the largest speedup gain of up to 247.5x for the TrueNorth Reference application.
In order to identify the optimal GPU thread block size that benefits the router implementation the most, we use the MNIST-12c, 128c and 512c, CIFAR-10, VMM 32x32 and TrueNorth Reference applications.  We observe that router implementation is not sensitive to change in thread block configuration at all. This minimal variation in speedup is reasonable since the router kernel implementation does not employ shared memory based optimization technique that is reliant upon GPU architecture.

We perform the same thread block size sweeping experiment for the scheduler kernel implementation as shown in Figure~\ref{fig:schedulerblocksize} using Cifar-10, MNIST-128c and VMM 32x32 as other three applications show no sensitivity to thread block size variation. Among these three applications though we note that thread block size of 128 results with better speedup than other configurations. Therefore we fix the thread block size as 128 for both the router and scheduler implementations when evaluating the end-to-end execution time of the GPU-RANC. 

\begin{figure}[t]
    \centering
    \hfill
    \includegraphics[width=0.9\linewidth]{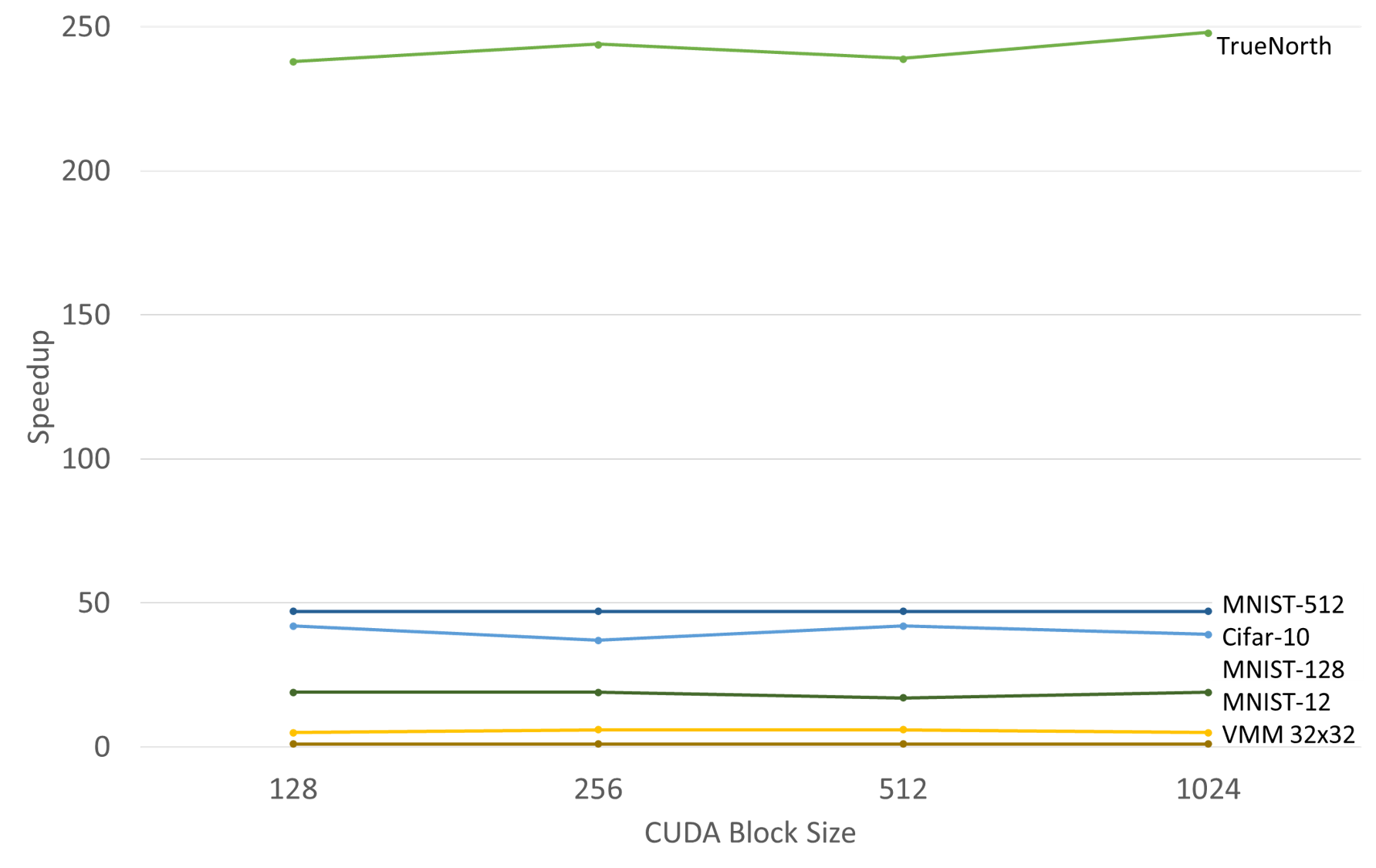}
    \vspace{-2pt}
    \caption{RANC Router Kernel Speedup vs CUDA Block Size. Speedup normalized against serial router time.}
    \vspace{-3mm}
    \label{fig:routerblocksize}
\end{figure}

\begin{figure}[t]
    \centering
    
    \hfill
    \includegraphics[width=0.9\linewidth]{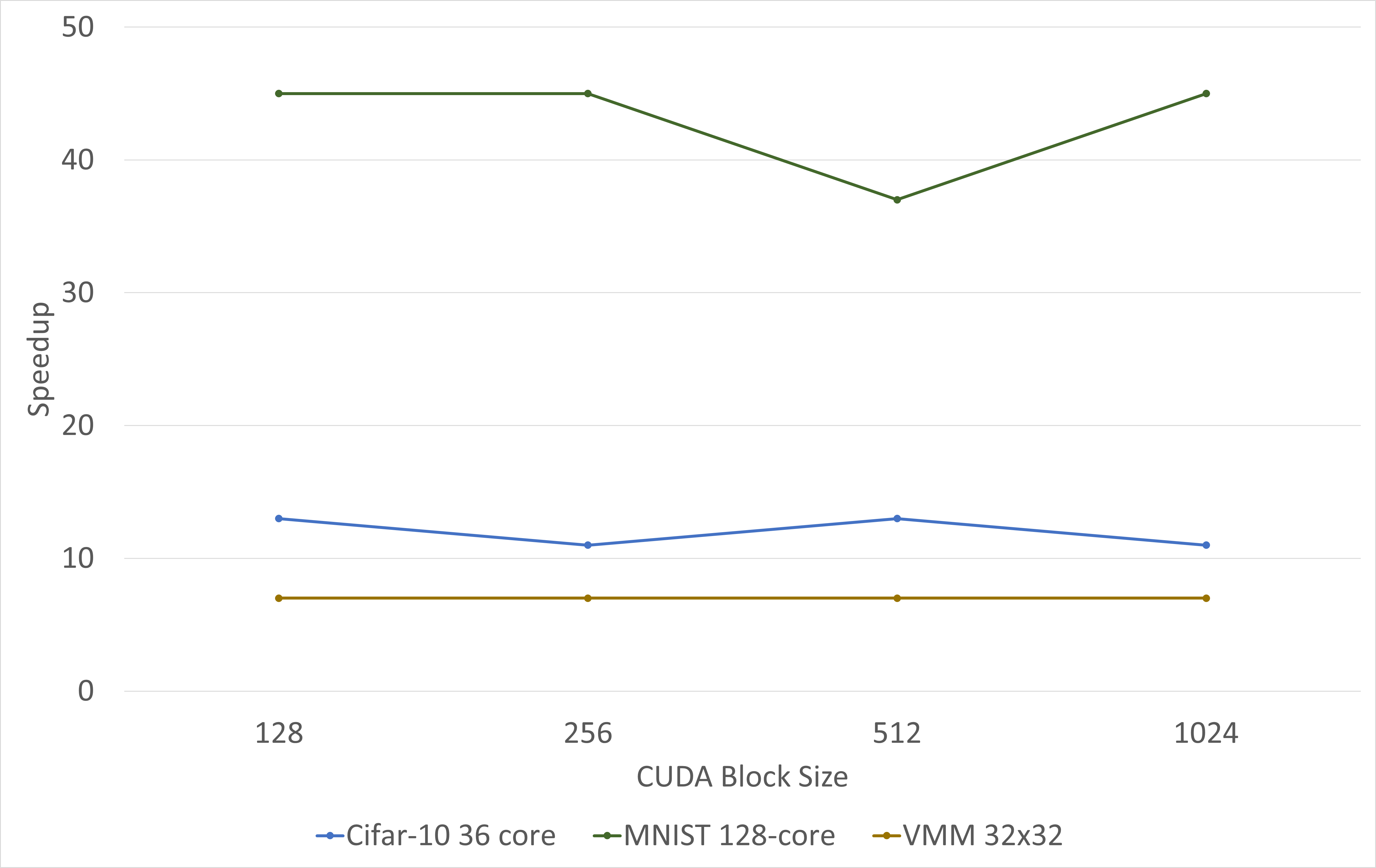}
    \vspace{-2pt}
    \caption{RANC Scheduler Kernel Speedup vs CUDA Block Size. Speedup normalized against serial  scheduler time.}
    \vspace{-5mm}
    \label{fig:schedulerblocksize}
\end{figure}

\begin{table}[t]
    \centering
    \caption{Scheduler Kernel: C++ vs. CUDA Performance}
    \begin{tabular}{|c|c c c|} 
        \hline
        Application & C++ Time (ms) & CUDA Time (ms) & Speedup Factor \\
        \hline
        MNIST-12c & 349 & 90.3 & 4.9 \\ 
        \hline
        MNIST-128c & 3706 & 81.7 & 45.4 \\ 
        \hline
        MNIST-512c & 13302 & 91.5 & 164.3\\ 
        \hline
        VMM 32x32 & 47 & 7 & 6.7 \\ 
        \hline
        CIFAR10 & 1069 & 80.3 & 13.3 \\ 
        \hline
        TrueNorth & 9103 & 10.6 & 858.7 \\
        \hline
    \end{tabular}
    \label{tab:schedulerresults}
     \vspace{-3mm}
\end{table}

Scheduler specific performance for all test runs is listed in Table~\ref{tab:schedulerresults} for thread block size of 128. Speedup ranges from 3.9x to 858.7x, with higher speedups mostly found in the longer running simulations. The longer the simulation, the more the initial GPU global memory allocation that occurs once at the simulator initialization becomes negligible. Total scheduler thread allocation scales with the number of axons and number of cores, so we see an uptick in speedup when scaling core counts within a fixed application such as MNIST. The final 512 core network has 4 times the core count of the 128 classifier, and we observe a speedup increase of 3.6x. This increase is slightly under the actual theoretical maximum increase likely due to flat overhead within the system such as allocating additional core SRAM data structures and CUDA memory copying. However, speedups across the board fall short of the level of parallelism that the kernel is acting upon (equivalent to the total grid axon count), which can be attributed to scheduler SRAM device allocation and data transfer penalties.

\begin{figure}[t]
    \centering
    
    \hfill
    \includegraphics[width=\linewidth]{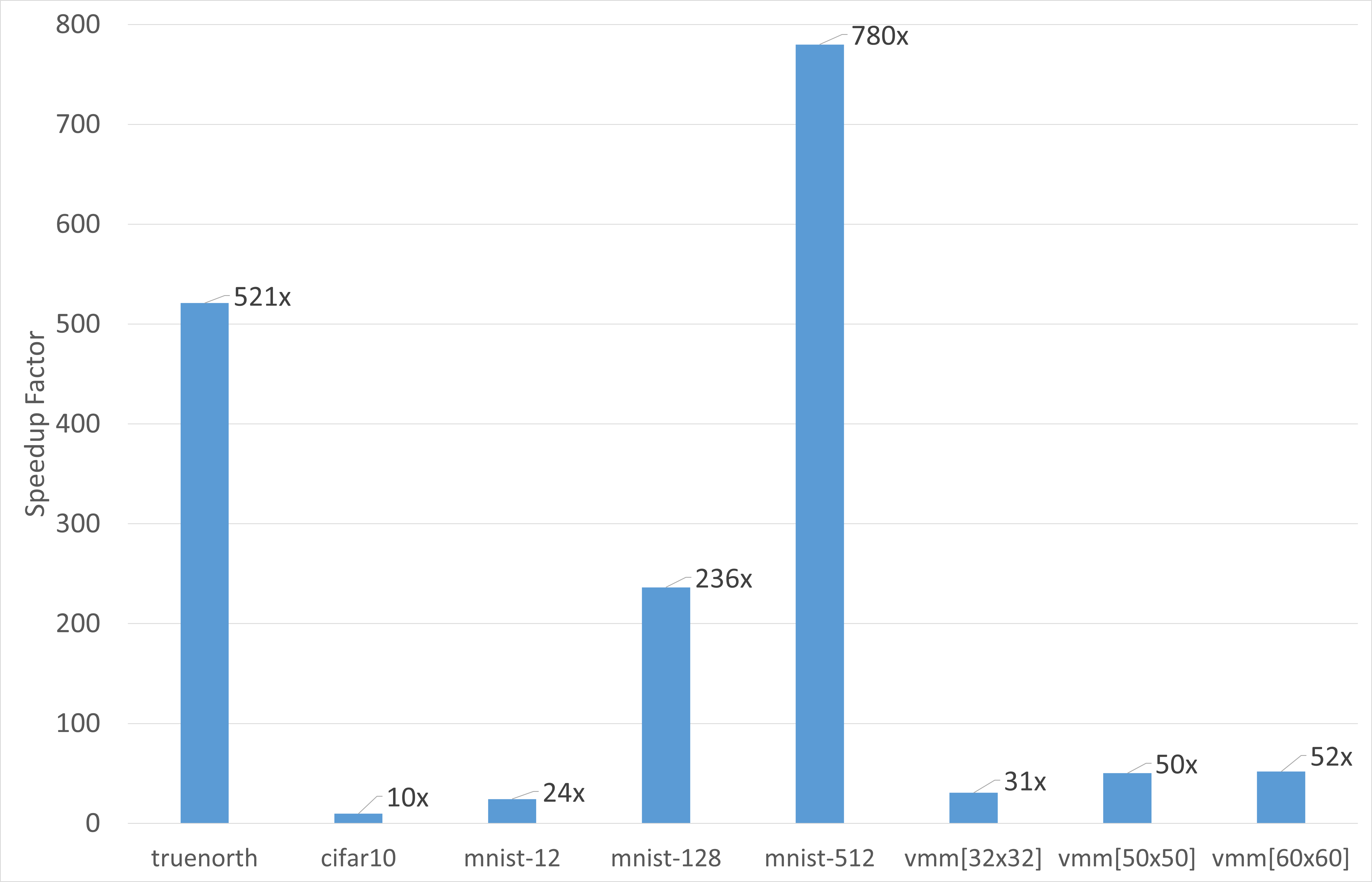}
    \vspace{-12pt}
    \caption{GPU-RANC end-to-end simulation time reduction amount with respect to the serial RANC for eight use cases.}
    \label{fig:allkernels}
    \vspace{-4mm}
\end{figure}
\subsection{End-to-End GPU-RANC Simulation Time Analysis}
In Figure~\ref{fig:allkernels} we present the execution time reduction amount with respect to the serial RANC simulator over 8 use cases for the final GPU-RANC configuration based on synaptic-level neuron block parallelization by launching each optimized kernel at the grid level within each tick. Referring to the execution time performance of the serial RANC for each application listed in Table~\ref{tab:applications}, we observe up to 780x reduction for the case of MNIST-512 core implementation scaling down the simulation time from 5.6 hours to 26 seconds.

Focusing on the MNIST application, speedup gain is 9.8x when core size increases from 12 to 128 and 3.3x when core size increases from 128 to 512. Ideal speedup gains would be 10.7x and 4x for these two scenarios respectively. The less than linear speedup gains are expected as we base our analysis on the end-to-end execution by taking time spent on data initialization, output generation, along with data transfer from CPU to GPU and from GPU back to CPU into account.    
For the different sizes of VMM application (32x32, 50x50, 60x60), the number of cores used in simulation ranges from 21 to 51. This makes all of the VMM applications to have less exploitable parallelism, resulting into speedup gains of 31x to 52x. Within the three sizes of the VMM, the 50x50 and 60x60 sizes require 45 and 51 cores respectively, and 512 axons both, which are much closer design sizes compared to the 21 cores with 256 axons/core needed for the 32x32 VMM. Hence, the speedup gain for the larger two VMMs are close in value.
TrueNorth Reference application, being the largest design with 4096 cores, experiences significant speedup benefit of 521x compared to the CPU-based simulation time. However, due to the significant increase in core count compared to the remaining applications, the kernel launches for this application also requires large number of GPU threads (134 million), leading to serialization of the operations by the GPU scheduler. Hence we observe comparatively less speedup than the MNIST-512c case.
From this result, we can conclude that several time consuming neuromorphic simulations, originally requiring in the magnitude of hours to complete, can now be completed in the magnitude of seconds with the benefit of GPU-RANC.

\section{CONCLUSION AND FUTURE WORK}
\label{sec:conclusions}
In this study we expand the capabilities of the open-source RANC ecosystem with a GPU-based simulation environment for performing fundamental research in applications and architectures of neuromorphic computing rapidly. RANC offers a flexible and highly parameterized environment for  hardware architects and application engineers to investigate and tune parameters of their neuromorphic architecture that would otherwise be unavailable on commercially available  purely prefabricated ASIC. The GPU-based RANC implementation offers ability to configure architectural parameters, including crossbar configuration in terms of the number of axons and neurons, the number of weights per neuron, and the bitwidths of weight, leak, neuron potential, and reset potential values. GPU-based parallelization  reduces the time scale of software-based simulations and paves the way for optimizing architectures based on application insights as well as studying hypothetical neuromorphic architectures rapidly. This in turn would allow faster convergence to the optimal hardware configuration and prototyping future neuromorphic architectures that can support new classes of applications entirely.  As future work our aim is to incorporate streaming-based execution flow on the GPU-RANC like the streaming supported with the FPGA-based RANC emulation environment. This would also lead to multi-GPU RANC implementation as a next step.

\bibliographystyle{IEEEtran}
\bibliography{refs.bib}

\end{document}